\documentclass[12pt]{iopart}
\usepackage{iopams}  
\usepackage{graphicx}
\usepackage{subfigure}
\usepackage{needspace}

\begin{document}
{\bf\Large{Trends in multiparticle production and some  ``predictions'' for pp and PbPb collisions at LHC.}}\\

\indent {\bf \large{Wit Busza}} \\
\indent \small{Massachusetts Institute of Technology, Cambridge MA USA}\\
\indent \small{E-mail: busza@mit.edu}\\

\begin{abstract}
Based on trends seen at lower energies we ``predict'' the multiplicities and pseudorapidity distributions of particle density and elliptic flow that will be seen in PbPb and pp collisions at the LHC.  We argue that, if these predictions turn out to be correct, either these quantities are insensitive to the state of matter created in high energy heavy ion collisions or the observed simplicity and universality of the data must be telling us something profound about the mechanism of particle production, which to this date is not well understood.
\end{abstract}


Two of the striking features of multiparticle production is simplicity and universality.  AA, pA, pp and even  $e^+e^-$  multiparticle production all exhibit trends, in particular energy and pseudorapidity dependence, that are remarkably similar and simple, and that do not follow in any obvious way from theoretical models.  I conclude from these facts that the currently accepted framework used in the interpretation of relativistic heavy ion phenomenology is missing some key ingredient  that is the root cause of this simplicity and universality.

In the not too distant future at the LHC, PbPb collisions will be studied at more than 25 times the highest energies available today, and pp at more than seven times.  If the trends seen to date continue at these higher energies, it will be more difficult than ever to dismiss the trends as accidental or unimportant.  It will become imperative that the origins of the simplicity and universality of the data be identified.  The reasons could be profound and give new insight into our understanding of ultrarelativistic heavy ion collisions and the state and interactions of the matter created in such collisions.  

With this in mind, in this lecture\footnote[7]{A summary of the results presented here was given at the ``Workshop on predictions in HIC at the LHC'', CERN, June 2007.  A more detailed version of this talk will be submitted for publication in Acta. Phys. Polonica.}, I summarize some important universal trends seen in AA, pA, pp and $e^+e^-$ charged particle production, integrated over all species and transverse momenta, and then use the trends to extrapolate to LHC energies the lower energy data.  Ie based on the data to date I ``predict'' some of the early results that will be seen at the LHC.

A key feature seen in the collision of all systems, be it  $e^+e^-$ , pp, pA, or AA, is that the dependence on energy of the pseudorapidity distributions does not seem to depend significantly on the nature of the incident systems, and the dependence on the incident systems of the pseudorapidity distributions does not seem to depend significantly on the energy of the collision.  For example, we find that in hadron-nucleus \cite{1} and AA collisions \cite{2} the total charged particle multiplicity is proportional to the total number of participants $N_{part}$ (wounded nucleons in the language of Bia{\l}as and Czy\.{z} \cite{3})with constant of proportionality independent of energy.  Furthermore we find that in the case of AA collisions, per participant pair, the total charged particle multiplicity is almost the same as in  $e^+e^-$  collisions at the same center of mass energy, $\sqrt{s_{NN}}$, per pair, and in the case of pA it is almost the same as in pp collisions \cite{2}.  As a result the $N_{part}$ dependence of the normalized total charged particle multiplicity produced in hadron-nucleus and nucleus-nucleus collisions can be summarized by a single universal curve, as illustrated in fig. 1.  It is remarkable that for such a variety of systems with $N_{part}$ ranging from 2 to 350, and energy from at least 10 GeV to 100 GeV, the $N_{part}$ dependence is the same.  Equally surprising is the observation that for all colliding systems the total charged particle multiplicity increases in a similar way, as $ln^2\sqrt{s_{NN}}$, with $\sqrt{s_{NN}}$ in GeV.  See fig. 2.

\begin{figure}[h] 
  \centering
  \subfigure[Universal $N_{part}$ scaling of the normalized total charged particle multiplicity in hadron-nucleus and nucleus-nucleus collisions for $\sqrt{s_{NN}}$ between 10 GeV and 200 GeV.  It is a compilation of E178 and Phobos data, taken from ref 4.  ]{
   \includegraphics[width=2.9in]{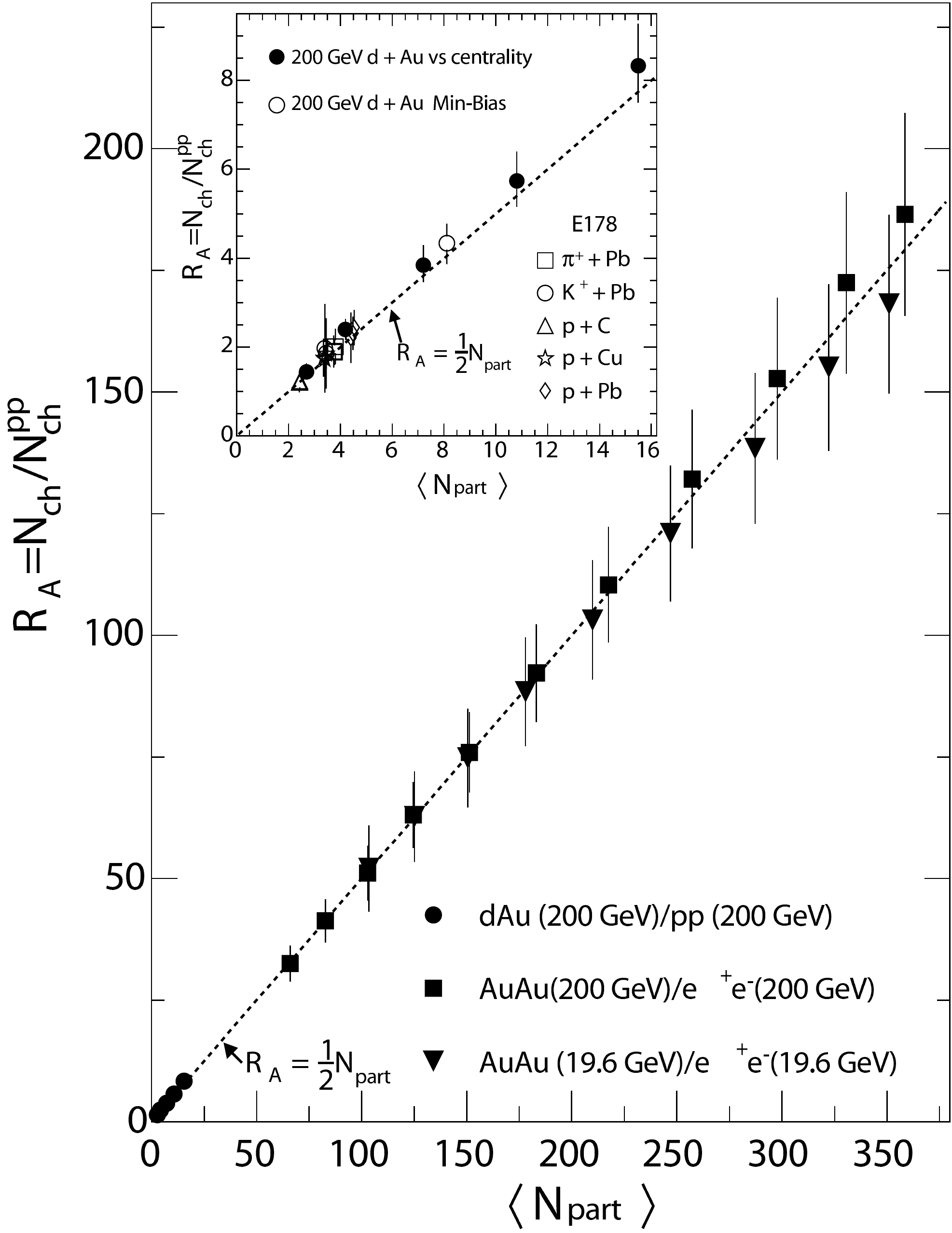}
   \label{fig1}}
 \hspace{0.05in}
 \subfigure[A compilation of data \cite{5} which shows that the total charged particle multiplicity in $e^+e^-$, pp, p\={p}, and AA collisions all scale with energy as $ln^2\sqrt{s}$, with $\sqrt{s}$ in GeV.]{

   \includegraphics[width=2.9in]{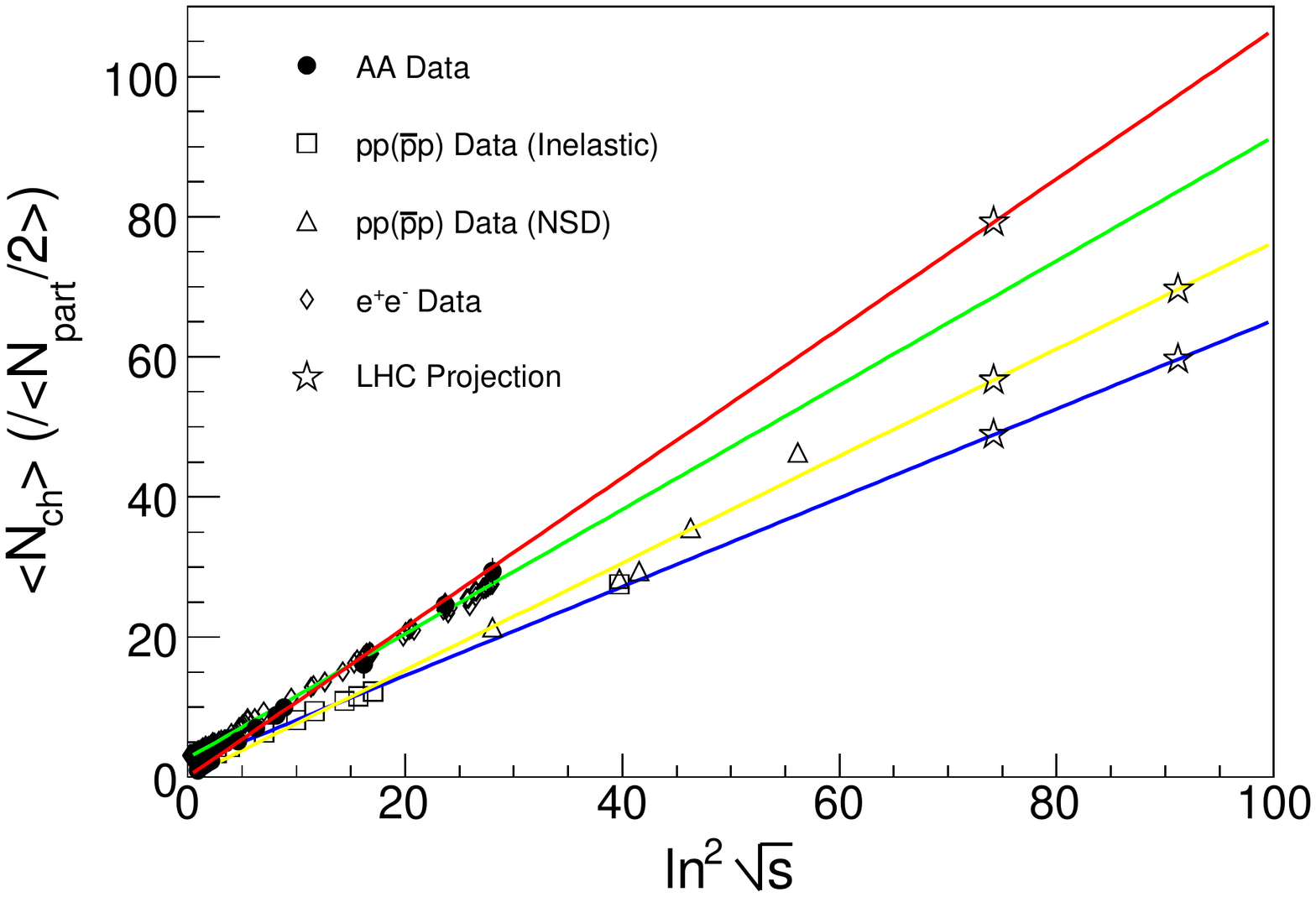}
  \label{fig2} }
  \end{figure}

Extrapolating the data shown in figs 1 and 2 to LHC energies, we ``predict'' the following for collisions at the LHC:

For non-single-diffractive (NSD) pp at $\sqrt{s}$=14 TeV, $n_{ch}=70\pm8$

For non-single-diffractive (NSD) pp at $\sqrt{s}$=5.5 TeV, $n_{ch}=57\pm7$

For inelastic pp at $\sqrt{s}$=14 TeV, $n_{ch}=60\pm10$

For inelastic pp at $\sqrt{s}$=5.5 TeV, $n_{ch}=49\pm8$

For PbPb collisions with $N_{part}$=386 (top 3\% centrality \cite{6}) at $\sqrt{s_{NN}}$=5.5 TeV, $n_{ch}=15000\pm1000$

\begin{figure}[h] 
  \centering
  \subfigure[Midrapidity particle density per participant pair for AA collisions.  Data are from a Phobos compilation \cite{2}.  Extrapolating the data to $\sqrt{s_{NN}}$ = 5.5 TeV yields  $\frac{dN_{ch}}{d\eta} =1200\pm100$ for $N_{part}$ = 386 (top 3\% centrality \cite{6}).]{
   \includegraphics[width=2.9in]{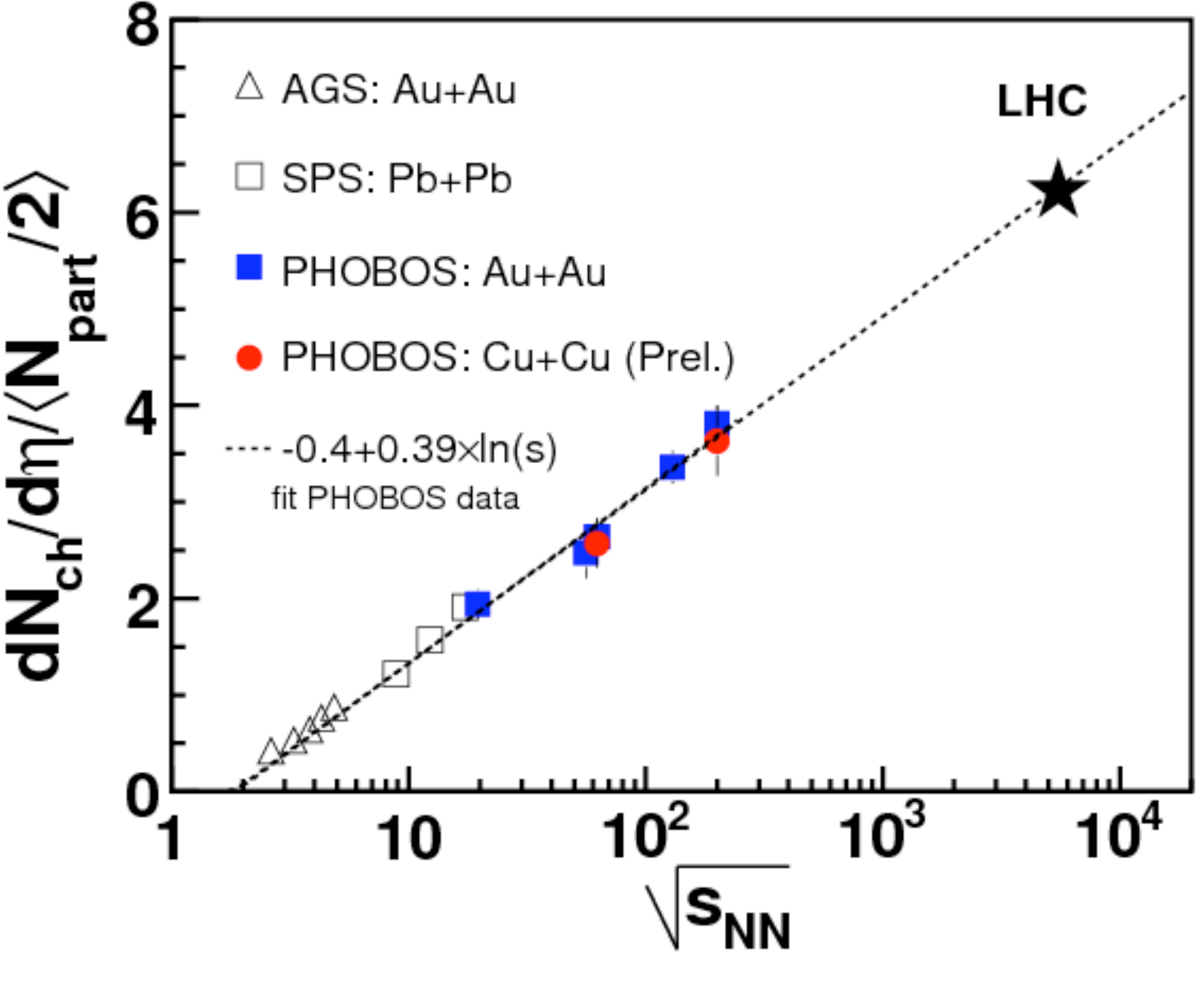}
   \label{fig3}}
 \hspace{0.05in}
 \subfigure[Examples of extended longitudinal scaling.  pp and p\={p} data \cite{8}, figure from ref 4; $e^+e^-$ data \cite{9}, figure from ref 2.]{
   \includegraphics[width=2.9in]{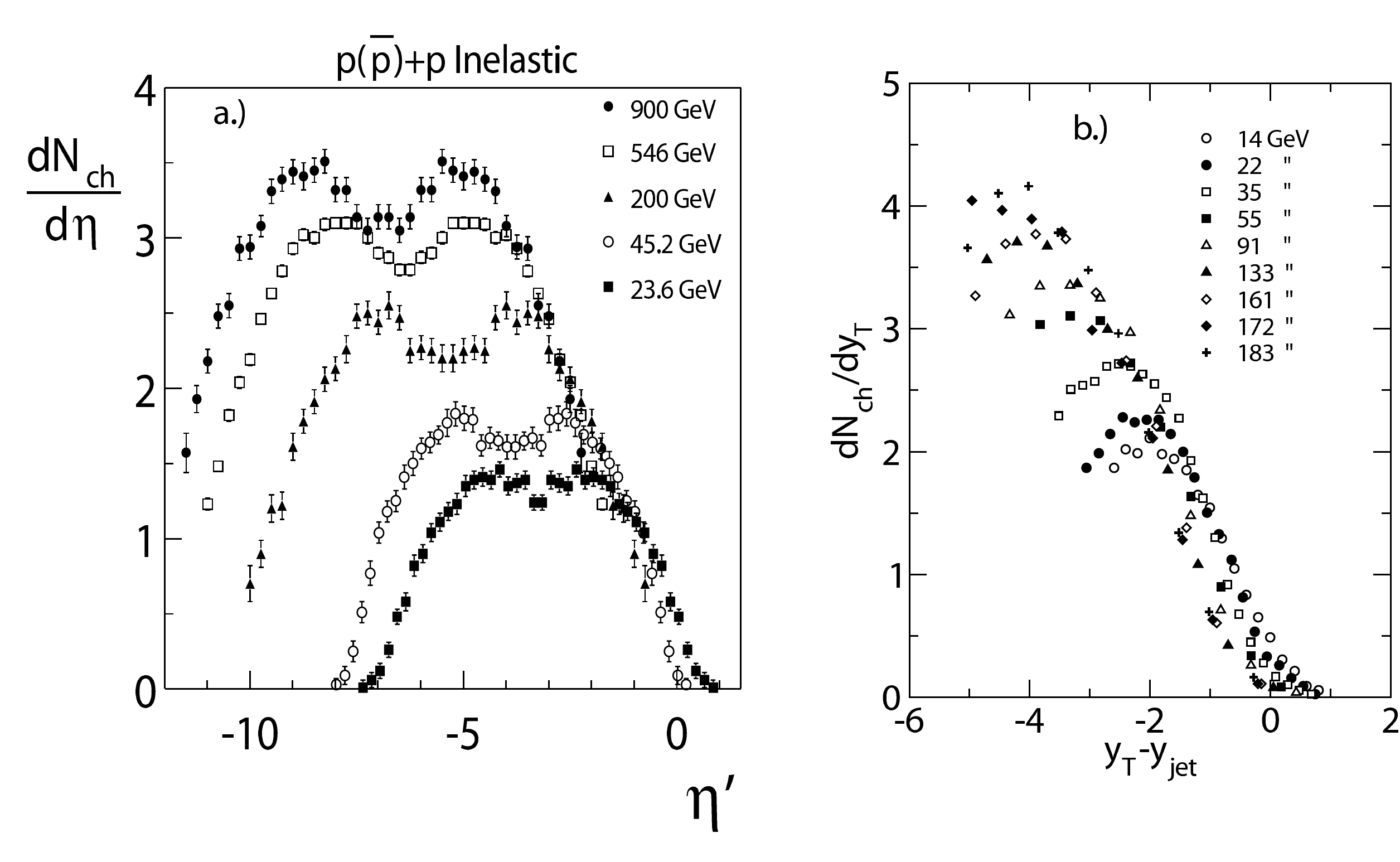}
  \label{fig4} }

\end{figure}

It turns out that the $ln^2\sqrt{s_{NN}}$ dependence of the total multiplicity is a direct consequence of three interesting features of the particle density pseudorapidity distributions: an increase as $ln\sqrt{s_{NN}}$ of the particle density at midrapidity, an almost trapezoidal shape of the  pseudorapidity distributions \cite{2}, and extended longitudinal scaling \cite{7} (a.k.a. limiting fragmentation) that leads to a growth of the width of the distributions proportional to $y_{beam} (\approx ln\sqrt{s_{NN}}$ at high energies).  An example of the $ln\sqrt{s_{NN}}$ growth of the midrapidity density for central collisions of heavy ions is shown in fig. 3.  Extended longitudinal scaling is most apparent if one plots the data in the rest frame  of one of the colliding systems, as shown for example in fig. 4 for pp and $e^+e^-$ collisions.

Using the observed energy dependence of the pseudorapidity distributions of the particle density described above, there are two mathematically equivalent ways in which we can extrapolate lower energy data to LHC energies.  The first method is illustrated in fig. 5 for central PbPb collisions.  By extrapolation of figs 1 and 2 to  $N_{part}$=360 (top 10\% centrality \cite{6}) and $\sqrt{s_{NN}}$ = 5.5 TeV we know approximately what will be the shape of the pseudorapidity distribution and its integral.  From the known $ln\sqrt{s_{NN}}$ growth of the midrapidity density for central heavy ion collisions (fig. 3) we know what will be the level of the plateau.  Finally we can use lower energy data and extended longitudinal scaling (ie. plot the lower energy data for central heavy ion collisions in the rest frame of the two colliding Pb nuclei at $\sqrt{s_{NN}}$= 5.5 TeV) to determine the slope and position of the sides of the ``trapezoid''.  All this information actually over constraints the ``predicted'' pseudorapidity distribution shown in fig. 5.

\begin{figure}[h]
 \centering
\subfigure[Extrapolation of Phobos AuAu data \cite{2} to $\sqrt{s_{NN}}$=5.5 TeV for PbPb with $N_{part}$=360 (top 10\% centrality) using extended longitudinal scaling and $ln\sqrt{s_{NN}}$ energy dependence of the midrapidity density.]{
   \includegraphics[width=2.9in]{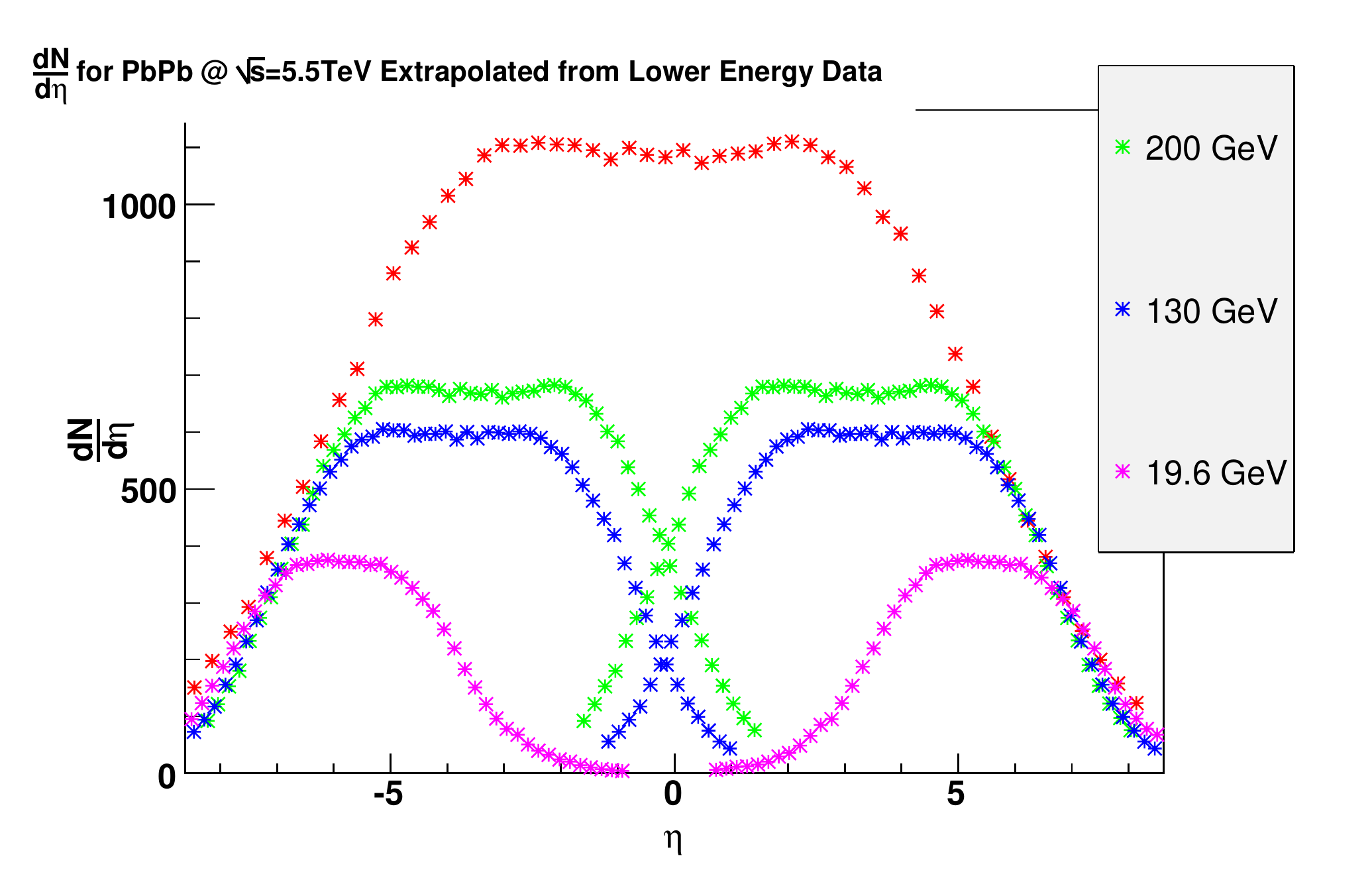}
  \label{fig5} }
\hspace{0.05in}
\subfigure[Extrapolation of Phobos AuAu \cite{2} and CuCu data \cite{10} for various values of $N_{part}$ to PbPb collisions with the same $N_{part}$ but $\sqrt{s_{NN}}$=5.5 TeV, using $ln\sqrt{s_{NN}}$ scaling of the width and height of the distribution (see text).  Note: Each curve is an independent extrapolation of a lower energy data set.]{
   \includegraphics[width=2.9in]{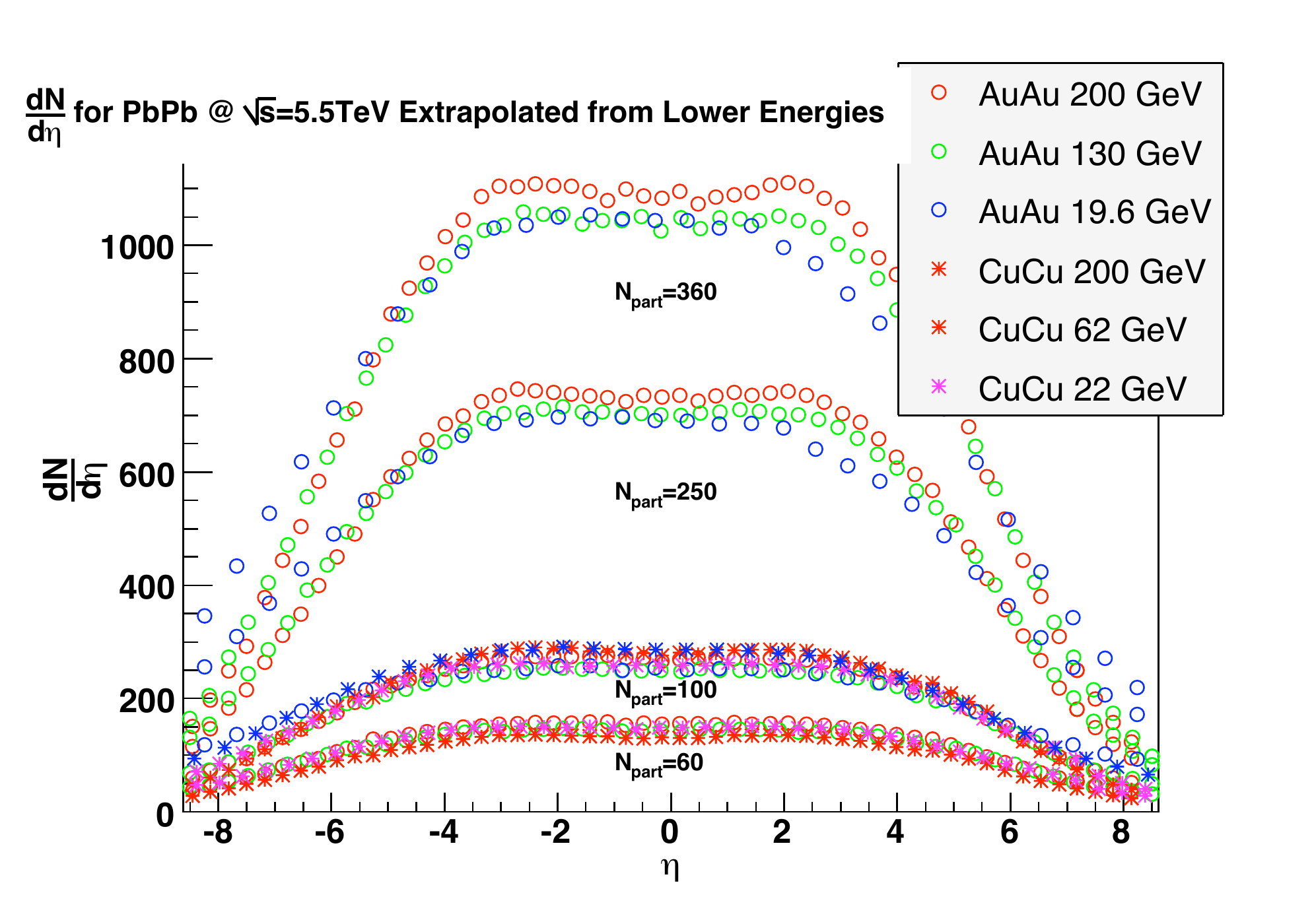}
  \label{fig6} }
\end{figure}

The second method uses the fact that the width and height of pseudorapidity distributions grow as $ln\sqrt{s_{NN}}$, that the shape of the distributions is approximately trapezoidal and that the particle density is essentially zero at beam rapidity.  As a consequence, for given colliding systems, the shape of the pseudorapidity distribution of the charged particle density is independent of energy, and both the width and the height are proportional to $ln\sqrt{s_{NN}}$.  Therefore any measurement of the distribution for some colliding systems at  one energy can be used to ``predict'' the result for the same colliding systems at any other energy .  In fig. 6 we use many data sets to ``predict'' LHC results for PbPb collisions with different impact parameter, and in figs 7 and 8 the same is repeated to predict pp and pA results respectively.  In the latter case, A is a hypothetical ``nuclear emulsion nucleus'' with effective $N_{part}$= 3.4.  The fact that the two methods, in fig. 5 and fig. 6, ``predict'' the same results and that in figs 6, 7, and 8 different data sets lead to the same ``prediction'' is direct evidence of the universal nature of the various trends seen in the data and discussed above.  [Note: the variation of the predicted results in fig. 7 most likely reflects the difficulty of reliable pp pseudorapidity measurements resulting from problems related to triggering]

\begin{figure}[h]
\centering
\subfigure[Extrapolation of pp Non-Single-Diffractive (NSD) data \cite{11} to $\sqrt{s}$=14 TeV using $ln\sqrt{s}$ scaling of the width and height of the distribution (see text).  Note: Each curve is an independent extrapolation of a lower energy data set.]{
   \includegraphics[width=2.9in]{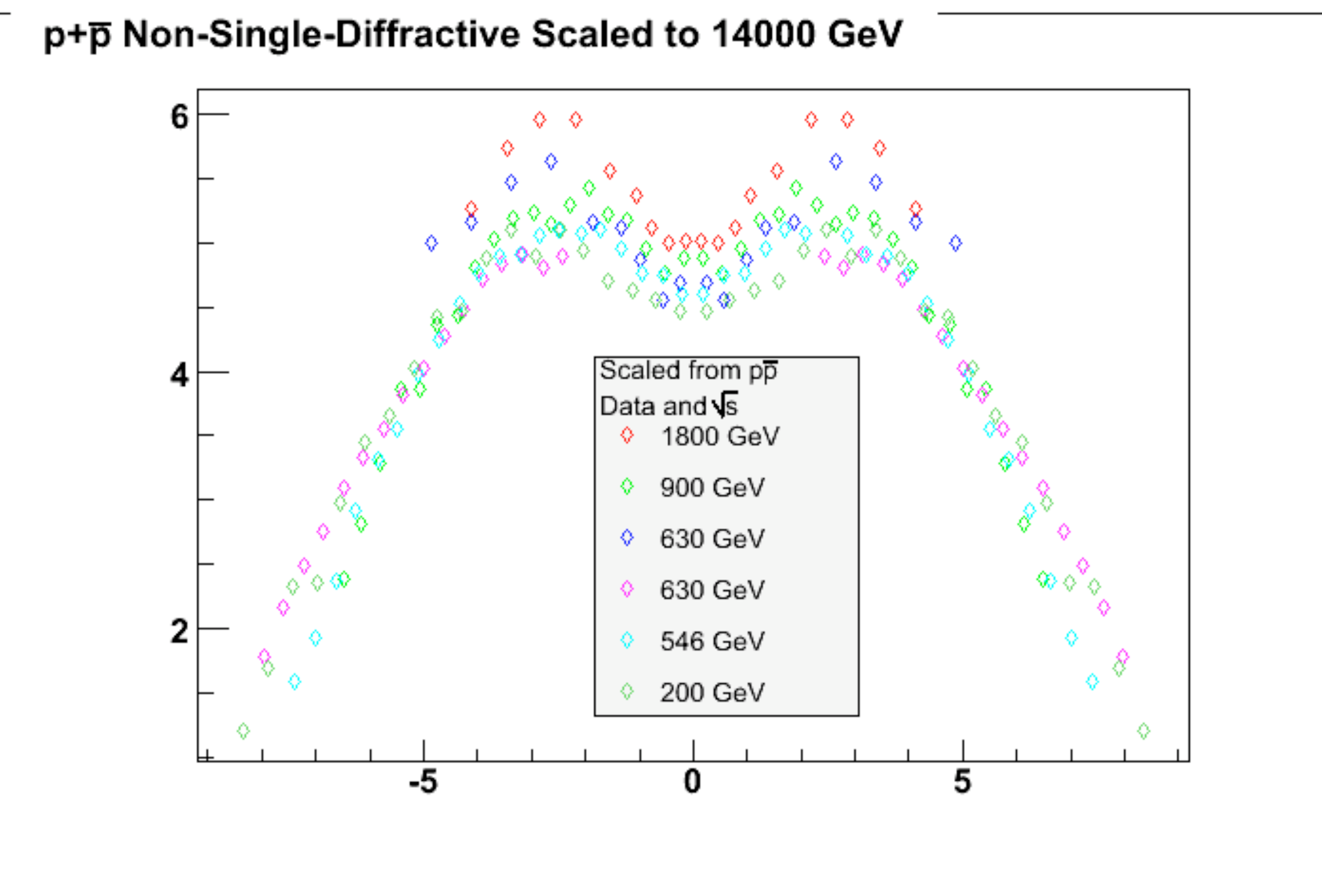}
  \label{fig7} }
\hspace{0.05in}
\subfigure[Extrapolation of p-Emulsion data to $\sqrt{s_{NN}}$=5.5 TeV for pA with $N_{part}$=3.4 (see text) using $ln\sqrt{s_{NN}}$ scaling of the width and height of the distribution.  Note: Each curve is an independent extrapolation of a lower energy data set.   The p-emulsion data \cite{12} are for proton beams with momentum 67 GeV/c to 800 GeV/c ($\sqrt{s_{NN}}$=11.3 GeV to 38.8 GeV).]{
   \includegraphics[width=2.9in]{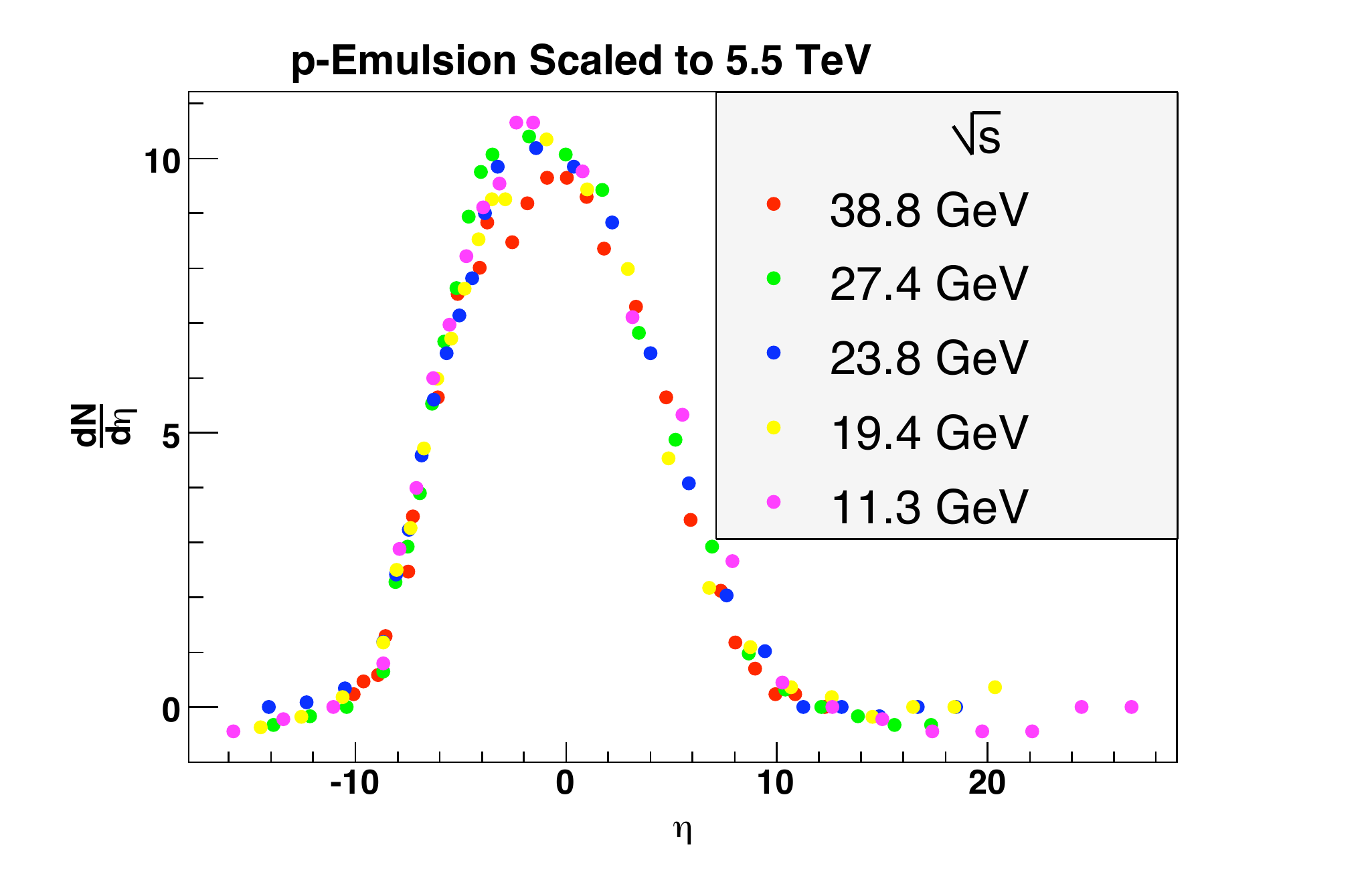}
  \label{fig8}}
  \end{figure}

As mentioned earlier, the total multiplicity in AA collisions depends only on $\sqrt{s_{NN}}$ and $N_{part}$.  However the detailed shape of the rapidity distributions depends on the geometry of the collision.  It is yet another interesting fact that at all energies and for different nuclei the shape of the distributions is almost the same provided the impact parameter corresponds to the same fractional cross-section (or better still the same value of $\frac{N_{part}}{2A} $ \cite{10}).  In figs 9 and 10 we use these facts to ``predict'' for PbPb collisions at LHC the centrality dependents of the shape of the pseudorapidity distributions and the midrapidity particle density.

\begin{figure}[h]
  \centering
\subfigure[Extrapolation of the shape of the particle density distributions, seen in Phobos AuAu data \cite{2}, to PbPb collisions at $\sqrt{s_{NN}}$=5.5 TeV for two centralities, 0-6\% and 35-45\%. See text.]{
   \includegraphics[width=2.9in]{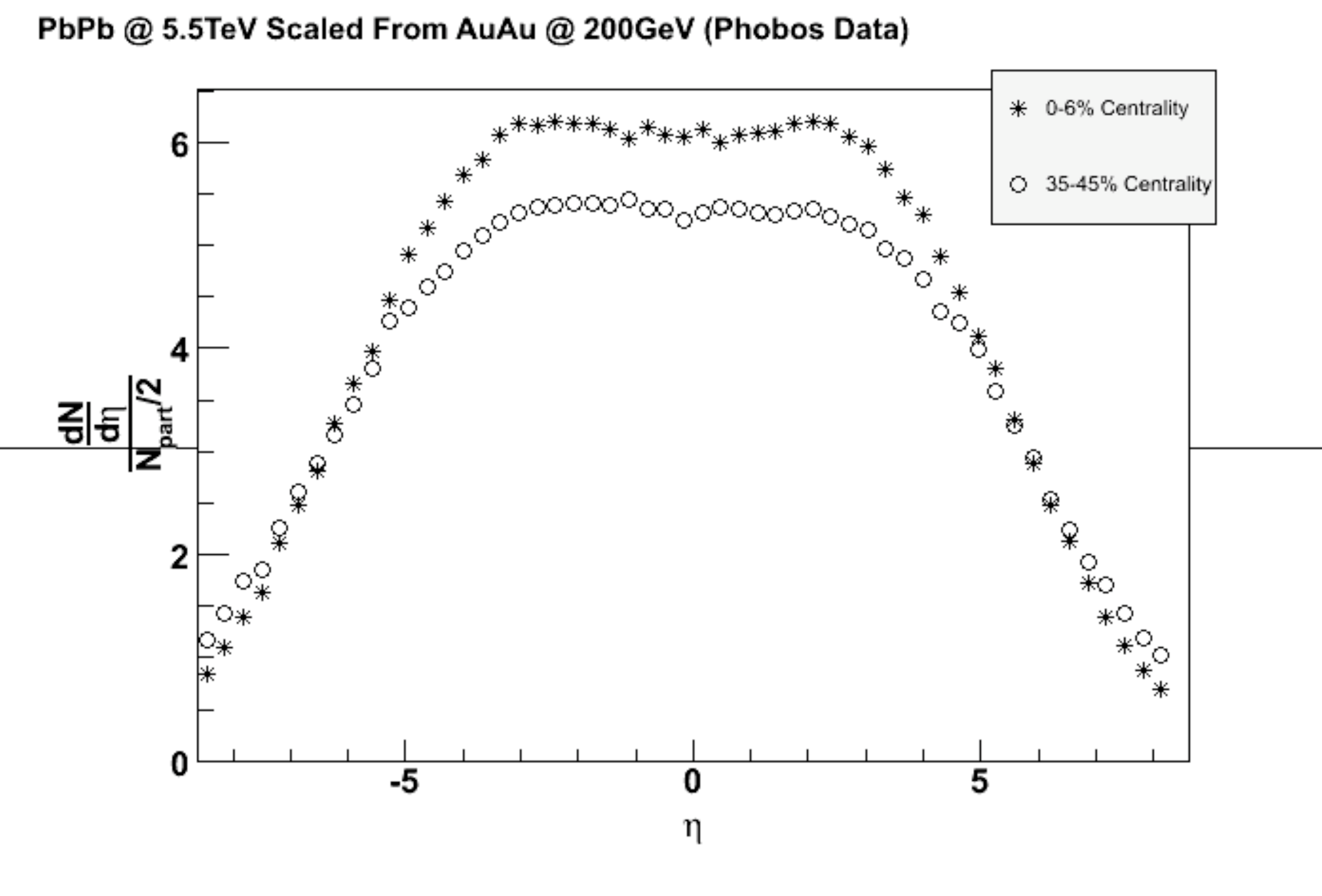}
  \label{fig9} }
\hspace{0.05in}
\subfigure[Extrapolation of the $N_{part}$  dependence of the midrapidity particle density seen in Phobos AuAu data \cite{2} to PbPb collisions at $\sqrt{s_{NN}}$=5.5 TeV.]{
   \includegraphics[width=2.9in]{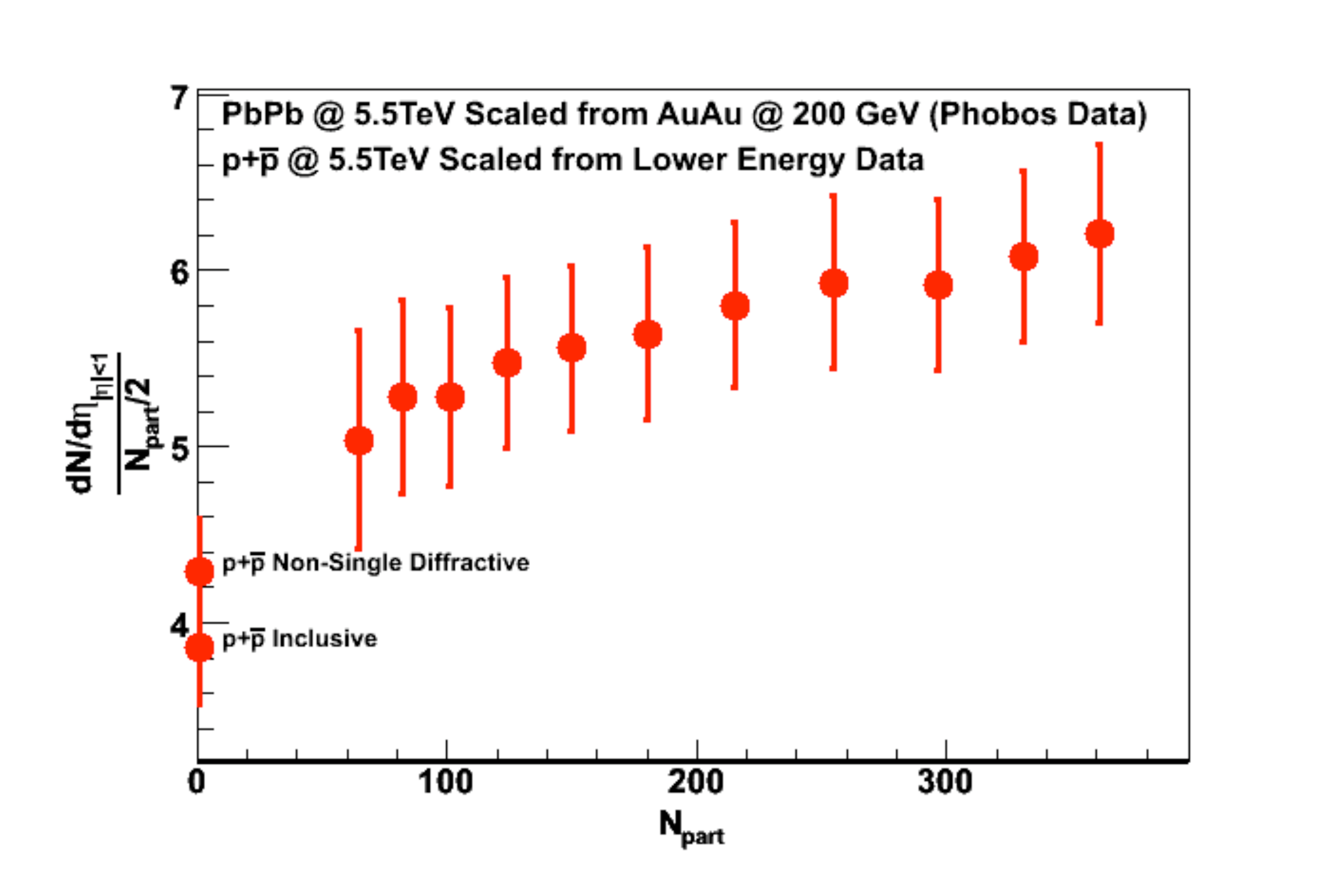}
  \label{fig10} }
  \end{figure}

Next, we briefly consider trends seen in the azimuthal distributions of particle density, or to be more specific, the pseudorapidity distributions of the elliptic flow parameter $v_2$ integrated over all charged particles and all transverse momenta.  Once again we find that for the same fractional cross-section at all energies, for both AuAu and CuCu  collisions, the shape of the rapidity distributions of the $v_2$ is the same, approximately triangular rather than the trapezoidal shape seen in the case of particle density.  Furthermore the energy dependence of the distributions again exhibit extended longitudinal scaling.  There is however one noticeable difference, the value of $v_2$ does not extrapolate to zero at beam rapidity.  At all energies the $v_2$ distributions appear to be on top of a ``pedestal''.  This may be an indication that the observed values of $v_2$ have an energy independent component, perhaps a non-flow contribution.  Whatever the reason for the ``pedestal'', as a consequence, we cannot use the second method to ``predict'' the $v_2$ distributions that will be seen at the LHC.  However we can still use the first method.  In fig. 11 the $v_2$ ``prediction'' is shown for the 40\% most central collisions of PbPb at $\sqrt{s_{NN}}$= 5.5 TeV.

\begin{figure}[h]
  \centering
\subfigure[Extrapolation of the elliptic flow parameter $v_2$ measured by Phobos at $\sqrt{s_{NN}}$ = 19.6, 62.4, 130.0 and 200 GeV for the 40\% most central AuAu collsions \cite{13}, to PbPb at $\sqrt{s_{NN}}$ = 5.5 TeV.  See text.]{
   \includegraphics[width=2.9in]{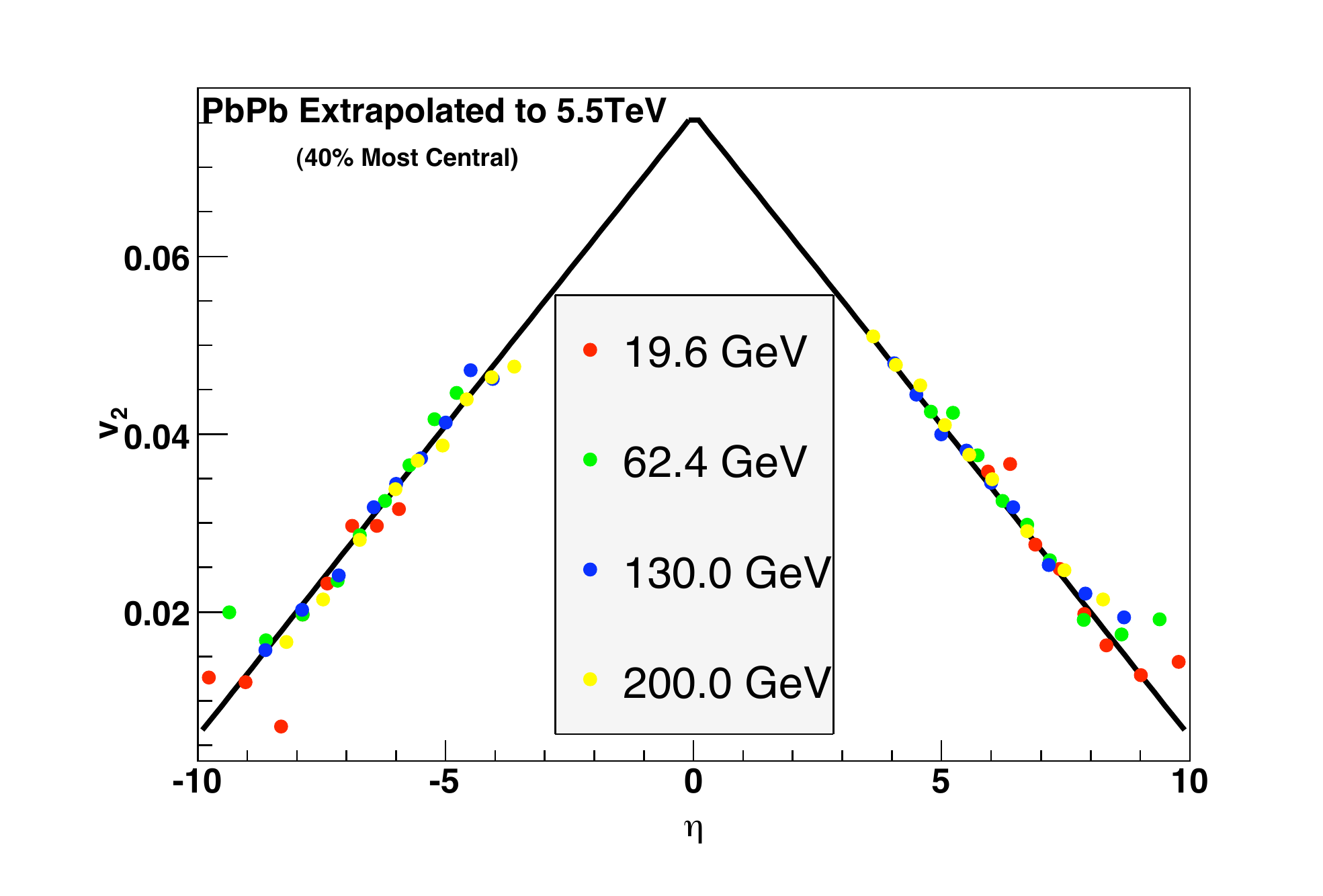}
  \label{fig11}} 
\end{figure}

So far I have tried to point out the remarkable simplicity and universality of multiparticle production, exhibiting such clear trends that it is possible to extrapolate the data to LHC energies with high precision.  I now briefly discuss why I find this data so intriguing.

In our current understanding, the processes taking place during particle production in $e^+e^-$ is quite different from that in pp collisions, let alone in AA collisions.  Even if we just consider AA collisions, our understanding is that there is a qualitatively different intermediate state produced in low and high energy AA collisions.  In the former the intermediate state is dominated by fragments of the incident nuclei, by baryons.  In the latter there is a relativistic hydrodynamic fluid of deconfined partons with extremely low viscosity.  How is it then that from $\sqrt{s_{NN}}$ below 10 GeV to above 200 GeV and for systems as diverse as $e^+e^-$, pp, pA and AA there are no obvious qualitative differences in the trends exhibited by the global multiparticle production data?  Or consider the difference in the pseudorapidity distributions of the particle density for peripheral and central AA collisions.  Per participant, the integral of the distributions is the same in the two cases, but not the detailed shape.  For central collisions there are more particles produced at midrapidity while for peripheral collisions there are more at high rapidity: an apparent exchange,  one for one, of slow particles and fast particles.  What is it in the mechanism of the particle production process that leads to an apparent conservation of particle number?  Surely these are not accidents!  There must be some general underlying physics, that to date is not well understood, that accounts for the observed simplicity and universality of the data.  LHC data, whether it is consistent or not with the extrapolations presented here, should throw much light on this topic.

As a final note, it is important to point out that if it turns out that, there are general principles that do show that the global properties of particles produced in high energy collisions are insensitive to the intermediate state, it follows that we can learn nothing about the intermediate state from the study of the global properties of multiparticle production, and furthermore that any successful theoretical model prediction of the global properties of the data cannot be used as evidence that the model correctly describes the intermediate state.

\newpage
I wish to acknowledge Alex Mott, Yen-Jie Lee, Andre Yoon, Richard Hollis and Rachid Noucier for help
with the plots.  This work was supported in part by DoE Grant DE-FG02-94ER40818.

References:

\end {document}